# Flexoelectricity in Amorphous Hafnium Oxide (HfO2)

*Daniel Moreno-Garcia[1], Kaitlin M. Howell[1], Luis Guillermo Villanueva[1]*

[1] Advanced NEMS Group, École Polytechnique Fédérale de Lausanne (EPFL), Lausanne 1015, Switzerland
E-mail: Guillermo.Villanueva@epfl.ch

**Abstract:**

Flexoelectricity, inherent in all materials, offers a promising alternative to piezoelectricity for nanoscale actuation and sensing. However, its widespread application faces significant challenges: differentiating flexoelectric effects from those of piezoelectricity and other phenomena, verifying its universality across all material structures and thicknesses, and establishing a comprehensive database of flexoelectric coefficients across different materials. This work introduces a groundbreaking methodology that accurately isolates flexoelectricity from piezoelectric, electrostrictive and electrostatic effects, with a detection threshold extending below 1 fC/m. The robustness of this method is demonstrated through its application to amorphous hafnium oxide (HfO2), successfully measuring a flexoelectric coefficient of 105 ± 10 pC/m. This measurement signifies the first measurement of flexoelectricity in hafnia, as well as in any amorphous material.
Additionally, the study compiles a list of published flexoelectric coefficients, revealing an important insight. The relationship between the flexoelectric coefficient and the material's relative permittivity is better approximated by a quadratic proportionality. This challenges the traditional linear assumption proposed in Kogan's work and opens new avenues for future research in flexoelectric materials.

**Introduction**

Flexoelectricity is a promising actuation and sensing technique when device dimensions reduce down to the nanoscale[1]. Its potential to outperform piezoelectricity in nanometer-scale applications[2] positions flexoelectricity as a possible protagonist for the future of nanoelectromechanical systems (NEMS)[3]. Unlike piezoelectricity, flexoelectricity is exhibited by all materials, irrespective of their symmetry group[4]. This universality eliminates the need for non-centrosymmetric materials. Additionally, flexoelectricity is not constrained by the Curie-temperature limitations that affect piezoelectric materials, allowing it to work at higher temperatures. Another notable drawback of some piezoelectric materials is their reliance on lead-based compounds, leading to toxicity concerns that restrict their use in many fields, including for example, biomedical applications[5]. In contrast, flexoelectric devices can be constructed from simple, non-toxic dielectrics.

Although nanoscale flexoelectricity offers promising characteristics, further efforts are needed to make it a practical technology. A key aspect involves creating a comprehensive and reliable database of flexoelectric coefficients for various materials while confirming that these coefficients remain constant at nanoscale thicknesses. Accomplishing these objectives requires a new method for measuring the flexoelectric coefficients of any material, ensuring they are distinctly identified and separated from other effects. This work aims to address this knowledge gap in the field.

First, this article reviews current methodologies for measuring flexoelectric coefficients. We highlight the specific challenges faced when measuring at the nanoscale and the difficulty in differentiating the flexoelectric effect from other phenomena such as piezoelectricity, electrostriction, and electrostatics. To overcome these, we propose a new measurement methodology that can accurately determine flexoelectric coefficients at the nanoscale and effectively isolate them from these rival effects. This methodology is applied to measure the flexoelectric coefficient of $HfO_2$ for the first time, representing a pioneering step in the measurement of flexoelectricity in a fully amorphous material.

The predominant method in the literature for measuring the flexoelectric coefficient involves inducing a deformation in the material and measuring the generated charges. These techniques rely on the so-called "direct" flexoelectric effect, where mechanical strain gradient generates polarization. The main distinction between different direct-flexoelectric measurement techniques lies in the method of strain (gradient) generation within the material. For instance, some studies use three-point bending testers, coupled with either a lock-in amplifier[6–9] or a charge amplifier and oscilloscope[10], to read the charges. Other experiments utilize a loudspeaker to produce the strain, with the charges measured using a lock-in amplifier. In these scenarios, the displacement of the loudspeaker is typically measured using a Displacement Voltage Ratio Transformer (DVRT)[11–14] or an optical sensor[15,16]. Additionally, some approaches utilize a piezo shaker to flex the sample, with the displacement controlled by a Laser Doppler Vibrometer (LDV). In these cases, charge measurements are conducted using either a lock-in amplifier[17] or a charge amplifier[18]. To ensure that the charges are not of piezoelectric origin, the experiments are often performed at a temperature above the Curie temperature.

The measurement of flexoelectric coefficients through the direct effect involves the detection of charges. As these charges tend to be small, the process is facilitated by using samples of large dimensions (millimeters to centimeters). This results in samples having thicknesses ranging from 0.5 to 3 mm, which aids in handling the sample and applying the load. Consequently, the methods described above are not ideal for measuring flexoelectricity at nanoscale thicknesses.

Measuring the flexoelectric coefficient of samples thinner than a few millimeters is possible through techniques utilizing the inverse flexoelectric effect, which involves observing the mechanical response of the material to an applied electric field. In such studies, micromechanical cantilevers are fabricated from the material being examined. When a voltage is applied to the devices, the resulting bending can be measured with a Digital Holographic

Microscope (DHM)[4] or a Laser Doppler Vibrometer (LDV)[19]. This technique allows for samples as thin as tens of nanometers to be measured.

The challenge in the above measurements lies in differentiating the flexoelectric effect from piezoelectric, electrostatic, and electrostrictive forces. In this paper, we present a methodology that represents an advancement over existing inverse flexoelectric measurement techniques[4], as it effectively isolates the flexoelectric effect from other concurrent phenomena (Figure 1a-d).

Our methodology is based on the measurement of the deflection of microcantilever beams composed of the material under test (in our case $HfO_2$) sandwiched between two metal layers (details on the fabrication are provided in the supplementary material). When applying a voltage across the dielectric layer (assuming the bottom electrode is grounded for simplicity) the cantilever bends due to one of the four phenomena mentioned above. The curvature of the beam due to flexoelectricity[20], as illustrated in Figure 1a, ($\kappa_{flexo}$) [1], is given by Equation 1. The effective flexoelectric coefficient is denoted by $\mu_{eff}$, the flexural rigidity is given by $D_f$ and $V$ is the applied voltage.

$$\kappa_{flexo} = \frac{\mu_{eff} V}{D_f} \quad (1)$$

However, there are still three other phenomena that can potentially create bending due to the horizontal expansion or contraction of the dielectric layer. To minimize their effect, the structure's neutral axis ($Z_{NA}$) should ideally be aligned with the center of the dielectric layer ($Z_D$). In other terms, the distance between the neutral axis and the center of the dielectric layer should be close to zero ($Z_p = Z_D - Z_{NA} \approx 0$). Unfortunately, fabrication uncertainties can offset the neutral axis, rendering this solution impractical. We must therefore consider that $Z_p$ might not be zero and that we might have another three additional effects.

The piezoelectric effect in the material, characterized by the coefficient $e_{31}$, causes the dielectric to elongate under the application of an electrical voltage (Figure 1b). Due to $Z_p$, this elongation induces a curvature[21], which is represented by Equation 2. More information on this formula can be found in the supplementary material.

$$\kappa_{piezo} = \frac{e_{31} Z_p V}{D_f} \quad (2)$$

The electrostatic force between the electrodes tends to pull them closer together when a voltage is applied, compressing the dielectric film positioned between them. Through the Poisson effect ($\nu_d$), this compression causes the film to expand in the perpendicular directions (Figure 1c). Given that the distance between the neutral axis and the center of the dielectric layer is likely not zero ($Z_p \not\approx 0$), this expansion creates a bending moment. The curvature resulting from this electrostatic effect involves vacuum and relative permittivity ($\varepsilon_0, \varepsilon_r$) and the dielectric thickness ($t_d$), as presented in Equation 3 and further explained in the supplementary material.

$$\kappa_{electrostatic} = -\frac{1}{2}\frac{\nu_d \varepsilon_r \varepsilon_0 Z_P V^2}{t_d D_f} \quad (3)$$

The electrostrictive effect is generated by the polarization of the material unit cell due to the applied electric field. This leads to an apparent piezoelectricity in the dielectric and a longitudinal elongation of the beam (Figure 1d). The curvature associated with this effect is given by Equation 4, and it depends on the Young's modulus of the dielectric ($E_Y$), and the electrostrictive coefficient $(M)$[22].

$$\kappa_{electrostrictive} = \frac{2E_Y M Z_P V^2}{t_d D_f} \quad (4)$$

Combining all these effects, the overall curvature as a function of voltage, $\kappa(V)$, is given by Equation 5.

$$\kappa(V) = \left(\frac{\mu_{eff}}{D_f} + \frac{e_{31} Z_p}{D_f}\right) V + \left(-\frac{1}{2}\frac{\nu_d \varepsilon_r \varepsilon_0 Z_P}{t_d D_f} + \frac{2E_Y M Z_P}{t_d D_f}\right) V^2 \quad (5)$$

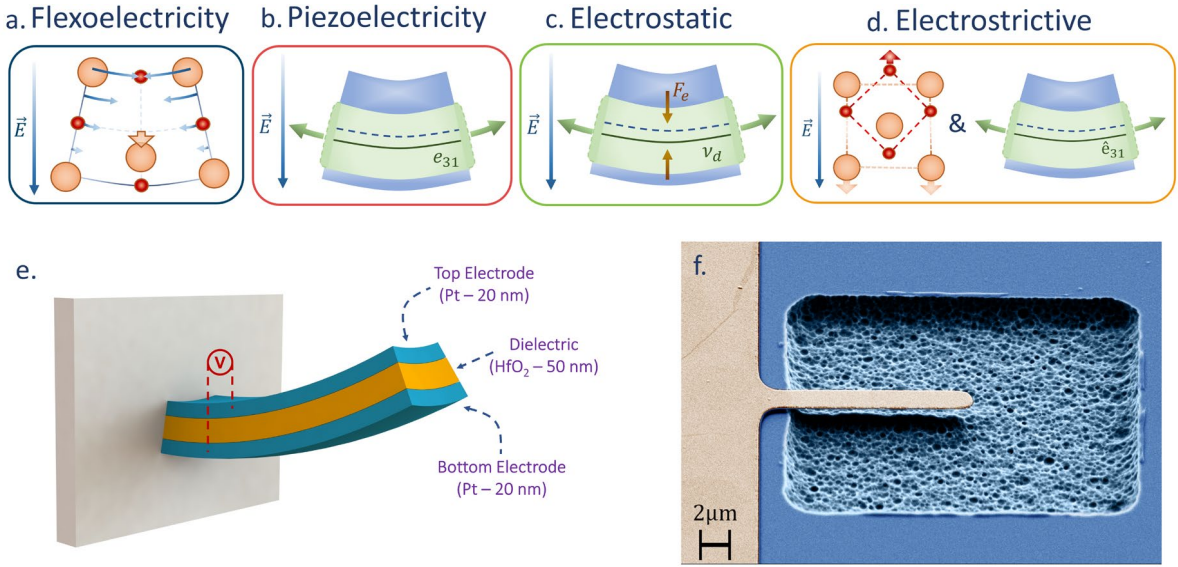

**Figure 1.** Different effects influence cantilever curvature under applied voltage. a) **Flexoelectric effect:** A strain gradient is generated in response to an electric field within the material, leading directly to beam curvature. b) **Piezoelectric effect:** The application of positive voltage elongates the dielectric (piezoelectric) material when the e31 coefficient is negative. This elongation induces curvature if the cantilever's neutral axis ($Z_{NA}$, dashed line in the schematic) is not aligned with the middle of the dielectric thickness ($Z_D$, continuous line in the schematic), i.e. $Z_p \neq 0$. c) **Electrostatic force:** This force attracts the two electrodes sandwiching the dielectric. As a result, the dielectric elongates due to the Poisson's ratio of the material. If $Z_p \neq 0$ it results in curvature. d) **Electrostrictive force:** The electric field polarizes the unit cell, causing the material to exhibit apparent piezoelectricity. If $Z_p \neq 0$ this leads to curvature. In the schematic, the electrostrictive force

coefficient ($M$) is assumed to have a negative sign. e) 3D representation of the cantilever, fabricated with Pt electrodes (20 nm) and the dielectric under study, in this case: HfO$_2$ (50 nm). f) SEM image of one of the fabricated cantilevers with a length of 14 µm and a width of 3 µm.

## Methods

The proposed methodology involves a two-step process: (i) Separate flexoelectricity and piezoelectricity from electrostatic and electrostrictive effects; (ii) Specifically isolating the flexoelectric effect by assessing the presence of piezoelectricity.

To achieve the first objective, we can simply actuate with a sinusoidal signal at a frequency ($\omega$). This approach allows for the differentiation between linear effects at $\omega$ (flexoelectric and piezoelectric) and the quadratic effects at $2\omega$ (electrostatic and electrostrictive). However, the potential presence of a residual DC component in the input complicates this differentiation. To account for it, measurements at different offset voltages are required[23]. However, instead of using different static DC values, which could suffer from time-drifting effects, more precise results can be obtained by applying a dynamic modulation to the actuation signal. In addition, our method also utilizes the resonant effect in the flexural devices under test, benefiting from the amplification associated with the quality factor. Two modulation approaches are employed:

$$\text{Amplitude Modulation:} \quad \sin(\omega_R t) \cdot (V_R + V_{mod} \cdot \sin(\omega_{mod} t))$$
$$\text{Two-tone Modulation:} \quad V_R \cdot \sin(\omega_R t) + V_{mod} \cdot \sin(\omega_{mod} t)$$

Here, $V_R$ represents the voltage at the resonance frequency $\omega_R$, while $V_{mod}$ denotes the voltage at the modulated signal frequency $\omega_{mod}$. The cantilever's displacement is measured using a Laser Doppler Vibrometer (LDV, Polytec), further details about the experimental setup are provided in the supplementary material. The displacement signal is then processed through a Lock-in Amplifier (Zurich Instruments, UHFLI), where it is demodulated at the resonance frequency ($\omega_R$), low-pass filtered, and we observe the resulting amplitude at $\omega_{mod}$ (after demodulation). By consistently performing these two experiments for each device, we can distinguish the effects that are linearly proportional to the actuation voltage – namely, flexoelectric and piezoelectric – from those that are quadratically proportional, such as electrostatic and electrostrictive forces. Table I presents the theoretical outcomes derived from this procedure.

| | Effects in Amplitude in cantilevers | |
|---|---|---|
| | Amplitude Modulation | Two-tone Modulation |
| Theoretical beam curvature at the probed frequency | $\kappa = \left(\dfrac{\mu_{eff}}{D_f} + \dfrac{e_{31}Z_p}{D_f}\right)V_{mod}$ | $\kappa = 2V_R V_{mod}\left(-\dfrac{v_d \varepsilon_r \varepsilon_0 Z_P}{2 t_d D_f} + \dfrac{2 E_Y M Z_P}{t_d D_f}\right)$ |
| Effects | $\sim \kappa_{flexo} + \kappa_{piezo}$ | $\sim \kappa_{electrostatic} + \kappa_{electrostrictive}$ |

**Table I.** Theoretical beam curvature resulting from the modulated signals applied to the cantilevers, followed by a two-demodulation step: first at $\omega_R$ and then at $\omega_{mod}$. The formulas in the table enable the separation of the contributions of flexoelectricity and piezoelectricity from those of electrostatic and electrostrictive forces.

The second step in the methodology involves distinguishing flexoelectricity from piezoelectricity. This distinction can be achieved based on their fundamental differences. As illustrated in Figure 1, piezoelectricity induces bending in the cantilever by causing elongation of the beam, while flexoelectricity results in pure bending due to a strain gradient without altering the length of the beam. This inherent difference in the nature of these actuation mechanisms has a very distinct effect on the resonance frequency of different flexural devices. In particular, piezoelectric actuation has been shown to cause much larger resonance frequency shifts on clamped-clamped structures rather than on cantilever beams[24]. On the other hand, flexoelectric actuation might induce lateral bending across the width of the beam, which could increase the stiffness and thus shift the resonance frequency in both types of structures[25,26].

Therefore, we can distinguish between piezoelectricity and flexoelectricity by observing changes in the resonance frequency when a two-tone modulation is applied. This is achieved through a phase lock loop (PLL, implemented using a Lock-in amplifier), which tracks and follows the shifts in resonance frequency. Table II shows the expected effects in frequency that flexoelectricity and piezoelectricity would produce in cantilevers and clamped-clamped beams.

|  | Flexoelectricity | Piezoelectricity |
|---|---|---|
| Cantilevers | $\Delta f_{Cant,Flexo}$ | $\Delta f_{Cant,Piezo}$ |
| Clamped-clamped beams | $\Delta f_{CC,Flexo}$ | $\Delta f_{CC,Piezo}$ |
| Expected effects | $\dfrac{\Delta f_{CC,Flexo}}{f_{CC}} \approx \dfrac{\Delta f_{Cant,Flexo}}{f_{cant}}$ | $\dfrac{\Delta f_{CC,Piezo}}{f_{CC}} \approx \alpha \cdot \dfrac{\Delta f_{Cant,Piezo}}{f_{Cant}}$ <br> For our geometry $\alpha \approx 200$ <br><br> $\dfrac{\Delta f_{CC,Piezo}}{f_{CC}} = \dfrac{e_{31} V}{2\sigma_0 t_{TOT}}$ |

**Table II.** Summary of the effects that induce changes in the resonance frequency of cantilevers and clamped-clamped beams when actuated with the two-tone modulation signal. Flexoelectricity causes changes in the resonance frequency of both cantilevers and clamped-clamped beams, which are similar in relative units. In contrast, piezoelectricity produces a larger change (by a factor of $\alpha$) in frequency for clamped-clamped beams compared to cantilevers. This factor is usually large and depends solely on the geometry and internal stress of the beam. For our geometry it has been simulated (COMSOL) to be $\alpha \approx 200$, further details are provided in the supplementary material. Additionally, the piezoelectric coefficient of the material can be calculated from the frequency shifts in the clamped-clamped beams, using the internal stress of the beam ($\sigma_0$), the total beam thickness ($t_{TOT}$) and the applied voltage (V).

Simulations performed in COMSOL, as detailed in the supplementary material, confirm that flexoelectricity induces a similar change in resonance frequency (expressed in relative units) in cantilevers and clamped-clamped beams (Table II). In contrast, for our geometry, piezoelectricity produces a change in resonance frequency that is approximately 200 times larger in clamped-clamped beams than in cantilevers[24] (Table II). Thanks to this significant amplification we can indeed compute the piezoelectric coefficient (or confirm that it is negligible), thereby distinguishing between flexoelectric and piezoelectric effects.

**Results**

We demonstrate the robustness of our methodology by experimentally determining the flexoelectricity in amorphous, undoped, hafnium oxide. To achieve this, we fabricate cantilevers and clamped-clamped beams composed of hafnium oxide with platinum electrodes. As an additional refinement to the methodology, the cantilevers are designed such that the neutral axis is close to the center of the beam, so that $Z_p \approx 0$ nominally, which helps minimize all three non-flexoelectric terms in Equation 5. The material under study, $HfO_2$, has a thickness of 50 nm and is sandwiched between two layers of platinum electrodes, each 20 nm thick (Figure 1e and 1f). The widths of all cantilevers and clamped-clamped devices are 3 µm, with cantilever lengths of 8, 10, 14, 16, 20 and 22 um, and clamped-clamped lengths of 60 and 80 µm. The amorphous state of the hafnia is confirmed through X-ray diffraction analysis, and the thicknesses are controlled with cross-section SEM images, details of which are included in the supplementary material.

The measurements follow the methodology detailed in "Methods" and use the set-up described in the supplementary material. Figures 2-4 show results from the intermediate steps to obtain the flexoelectric coefficient. Figure 2a shows the displacement of the tip of a 14-µm long cantilever when actuated by an amplitude modulation signal. We observe a linear dependence of the displacement with the modulation voltage ($V_{mod}$) while remaining invariant to the voltage at the resonance ($V_R$). This behavior aligns with the expectations outlined in Table I. The inset in Figure 2a shows the slopes of the displacement data, which emphasizes that they are constant for different $V_R$. This experiment provides the magnitude of the combination of flexoelectric and piezoelectric forces in the device.

Similarly, Figure 2b demonstrates the application of the two-tone modulation to the same cantilever. Here, the amplitude of the cantilever's response depends linearly on both $V_R$ and $V_{mod}$, aligning with the theoretical predictions of Table I. The inset in Figure 2b highlights the linear dependence of the response curve slopes on $V_R$. This experiment tells us the magnitude of the combination of electrostatic and electrostrictive forces.

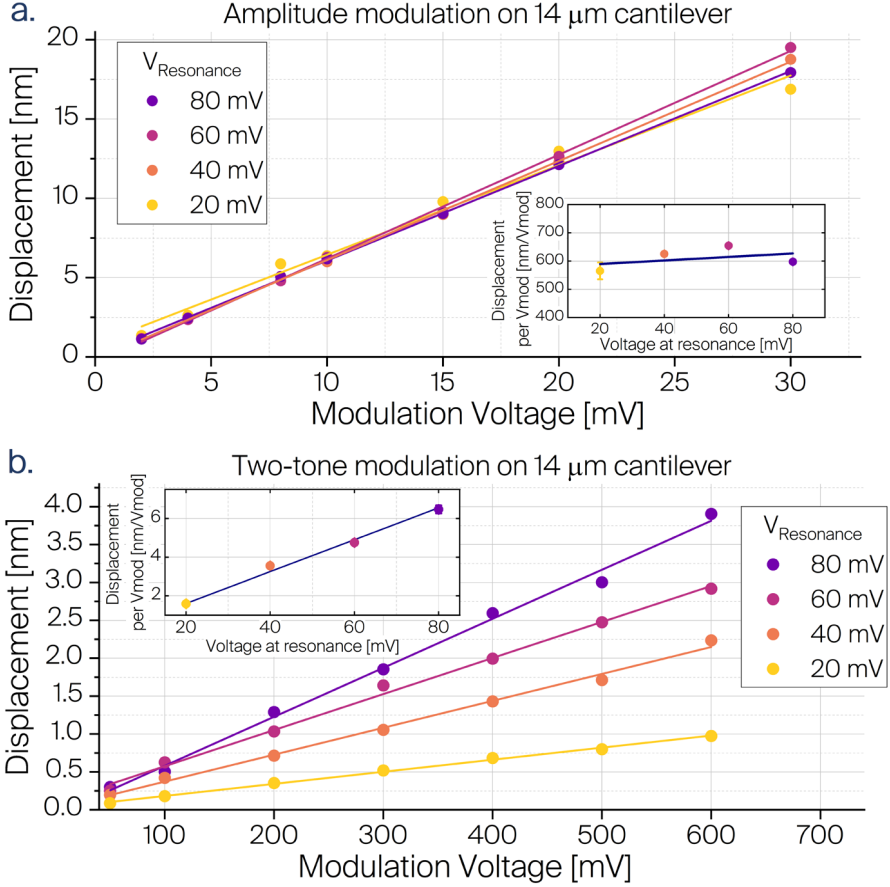

**Figure 2.** Separation of flexoelectric and piezoelectric responses from quadratic effects in a 14-μm long cantilever. (a) Displacement of the cantilever tip under the amplitude-modulated signal, showing linear dependence on modulation voltage ($V_{mod}$) and invariance to resonance voltage ($V_R$), consistent with Table I predictions. Inset: Slopes from the displacement data, quantifying flexoelectric and piezoelectric forces on the cantilevers. (b) Response of the same cantilever to the two-tone modulation signal, revealing linear dependence of response amplitude on both $V_R$ and $V_{mod}$, aligning with the theoretical expectations of Table I. Inset: Slope of each voltage sweep, showing the impact of electrostatic and electrostrictive forces.

In a series of extended experiments, cantilevers of varying lengths are measured following the same method. Figure 3a illustrates that longer cantilevers experience greater tip displacements at the same modulation voltages. This observation is attributed to the fact that, with identical curvatures, longer cantilevers naturally have larger tip displacements.

Figure 3b compiles the results obtained from these cantilevers when subjected to a two-tone modulation signal. This shows their dependence on the resonance frequency voltage $V_R$, as was theoretically anticipated in Table I. The comparison between Figure 3a and 3b reveals that the effect of piezoelectricity and flexoelectricity is significantly higher – by two orders of magnitude – compared to the effects of electrostatic and electrostrictive.

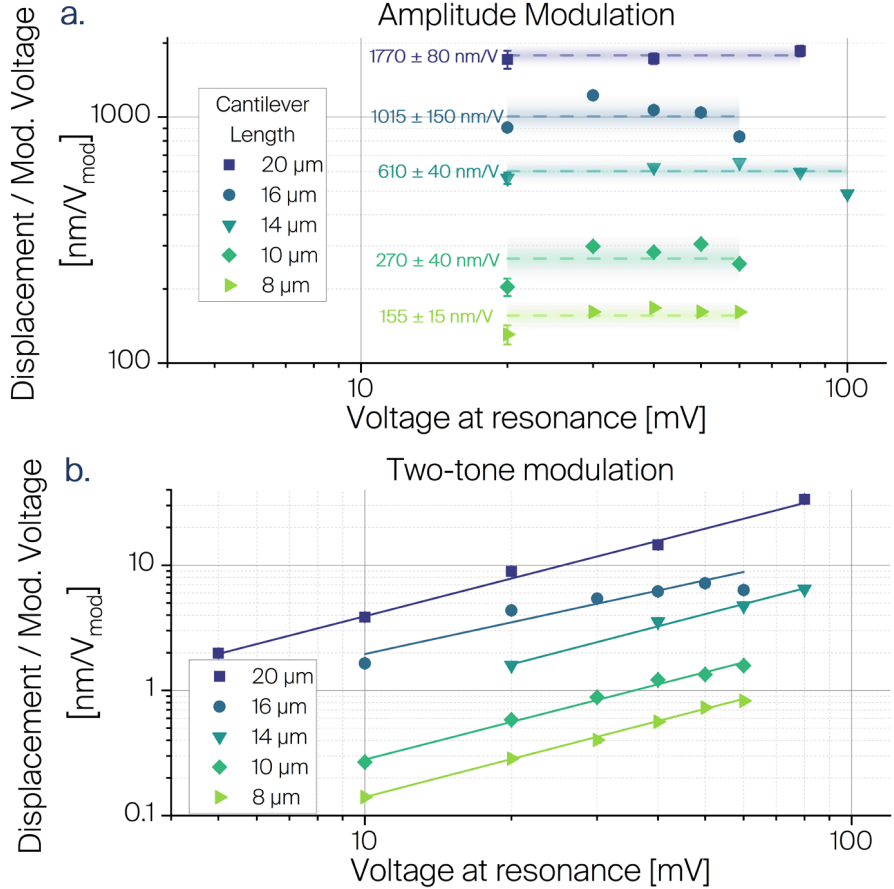

**Figure 3:** Separation of flexoelectric and piezoelectric responses from quadratic effects in a cantilever of various lengths. (a) Effect of the amplitude modulation signal applied on cantilevers. Shows that longer cantilevers exhibit greater tip displacement at identical modulation voltages. This experiment gives the contribution of flexoelectricity and piezoelectricity of the cantilevers. (b) Compilation results from cantilevers of different lengths under the two-tone modulation signal, offering insights into the electrostatic and electrostrictive effects. The comparison between (a) and (b) highlights the higher impact of piezoelectricity and flexoelectricity, which exceeds the electrostatic and electrostrictive effects by two orders of magnitude.

After successfully completing the first step in our methodology, isolating the contributions of flexoelectricity and piezoelectricity, the subsequent phase distinguishes between these two effects. This differentiation is achieved by analyzing the shifts in resonance frequency observed in both cantilevers and clamped-clamped beams. The effect on clamped-clamped beams is expected to be larger[24], it will depend on the geometry of the beams and can be calculated with finite element simulations (see supplementary material). According to our COMSOL simulations piezoelectricity in the material is expected to cause a frequency shift that is 200 times more pronounced in clamped-clamped beams than in cantilevers, for our geometry. This significant amplification serves two primary purposes: (i) it acts as a definitive indicator of piezoelectricity within the material, and (ii) if piezoelectricity is present, it allows for a more straightforward calculation of the piezoelectric coefficient (Table II).

In our experiments, we observe similar resonance frequency shifts in both cantilevers and clamped-clamped beams. Figure 4 presents an example to illustrate this fact. Specifically,

Figure 4a details the effects observed in a 20 μm cantilever, while Figure 4b examines a 60 μm clamped-clamped beam. This observation effectively rules out the presence of the piezoelectric effect. The origin of the observed shifts in frequency therefore points to the flexoelectric effect. One hypothesis, which requires further experiments, is that flexoelectricity also bends the cantilevers laterally (along the width of the structures), which causes stiffening and changes in the resonance frequency.

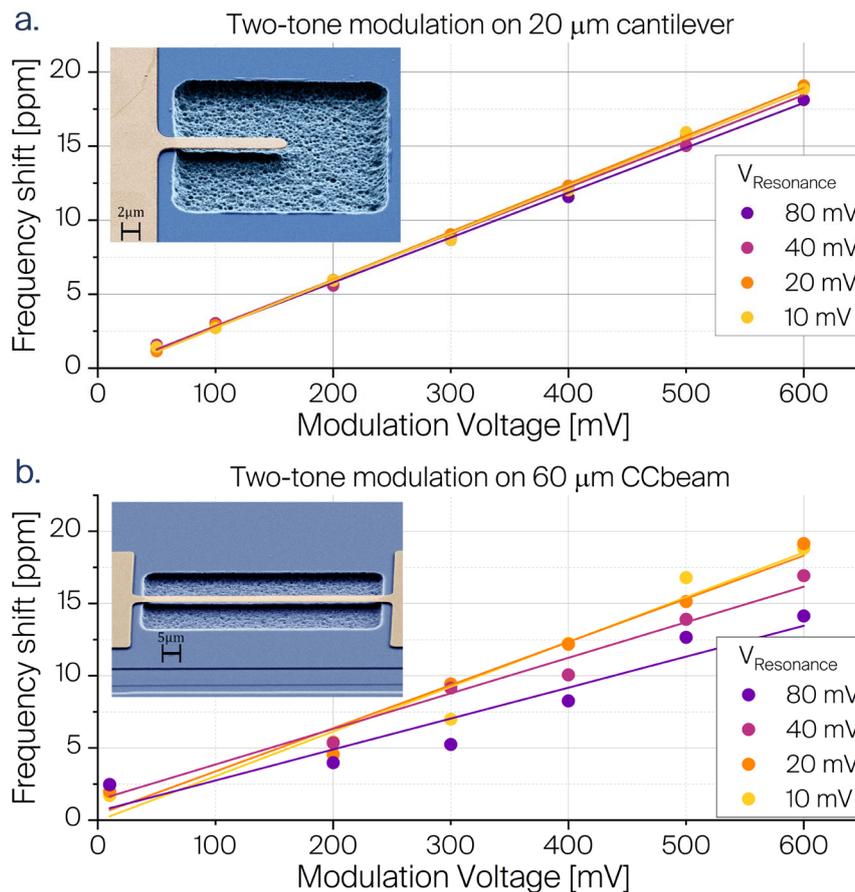

**Figure 4.** Discriminating flexoelectric and piezoelectric effects. Comparative analysis of resonance frequency shifts in cantilever and clamped-clamped beam. (a) Displays resonance frequency shifts in a 20 μm cantilever. (b) Shows resonance shifts in a 60 μm clamped-clamped beam. The similarity in the frequency shifts observed in both structures contradicts the expected 200-fold increase in the clamped-clamped beam, suggesting a negligible piezoelectric influence in the material.

Once the absence of the piezoelectric effect in our samples is confirmed, we use the measurements presented in Figure 3a to calculate the flexoelectric coefficient. This involves converting the displacement at the tip into cantilever curvature (see supplementary material) and subsequently extracting the flexoelectric coefficient from Equation 1. We perform these calculations across devices of varying lengths, observing that the flexoelectric coefficient remained consistent. Figure 5 provides a comprehensive summary of these values extracted from the different measured cantilevers. We determine the mean flexoelectric coefficient for (our) amorphous $HfO_2$ material to be $105 \pm 10$ pC/m.

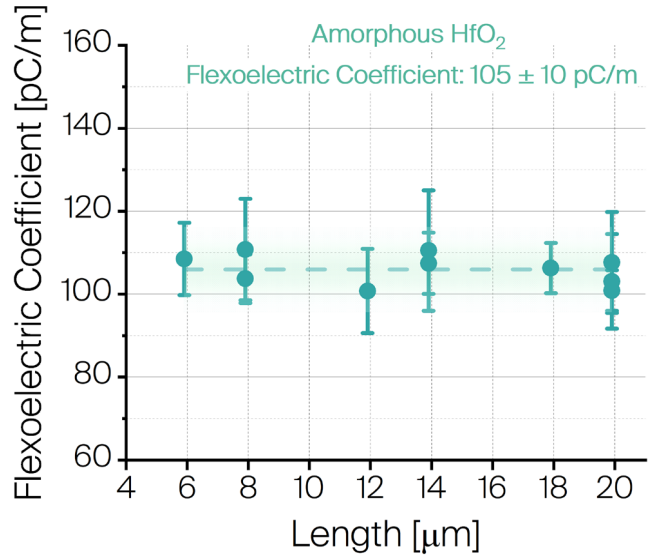

**Figure 5.** Flexoelectric coefficient measurements across varied cantilever lengths. The data indicates consistent flexoelectric coefficients irrespective of cantilever length. Based on these findings, the flexoelectric coefficient of $HfO_2$ is determined to be $105 \pm 10$ pC/m. This not only represents the first measurement of flexoelectricity in $HfO_2$ but also marks the first demonstration of flexoelectricity in any amorphous material.

In this study, we report the smallest flexoelectric coefficient ever measured. Additionally, our methodology is capable of measuring coefficients as low as $1\,fC/m$. Detailed explanations on the detection limit are provided in the supplementary material, but essentially, the primary limitation of our methodology is the thermomechanical noise inherent in the devices. Consequently, this establishes our protocol as a superior technique for precise measurements of flexoelectricity.

**Discussion**

According to theoretical estimates, the magnitude of flexoelectricity is believed to linearly depend on the material's susceptibility ($\chi$) or relative permittivity[1,20,27–29]. The proportionality constant was originally predicted by Kogan[28] and established an upper bound to the flexocoupling coefficient ($f \equiv \mu_{flexo}/\chi = 1 - 10\,V$) in simple ionic solids, a range now referred to as "Kogan's estimate".

However, literature presents numerous instances where the proportionality coefficient significantly exceeds Kogan's estimate[8,10,11,14,16,30], reaching several hundred volts for some flexocoupling coefficients. These unexpected findings have prompted investigations to explain the anomalously large (often called "giant") flexoelectric coefficients. These have pointed towards surface piezoelectricity[8,31], presence of Schottky barriers on the metal-dielectric interface[32], nano-[30] or micropolar regions[33], strong lattice instability[10], processing-induced strain gradients[34] or remnant piezoelectricity above the transition temperature[33].

To compare our results with those in the literature, we present Figure 6, which compiles a comprehensive dataset of flexoelectric coefficients from various materials and compares them as a function of their relative permittivity. Based on the literature analysis and contrary to Kogan's theoretical predictions, a quadratic relationship better fits the relationship between the flexoelectric coefficients and relative permittivity. Notably, works previously labeled as exhibiting 'giant' flexoelectric coefficients[10,16] align well with this quadratic trend. At this point, it is premature for us to hypothesize the reasons for this difference in behavior. Additionally, we cannot discard the possibility of piezoelectric contributions affecting certain cases. Previous methods for measuring flexoelectricity may not have clearly differentiated it from piezoelectric effects[35]. In contrast, our method makes sure that the flexoelectric coefficients do not have a piezoelectric contribution.

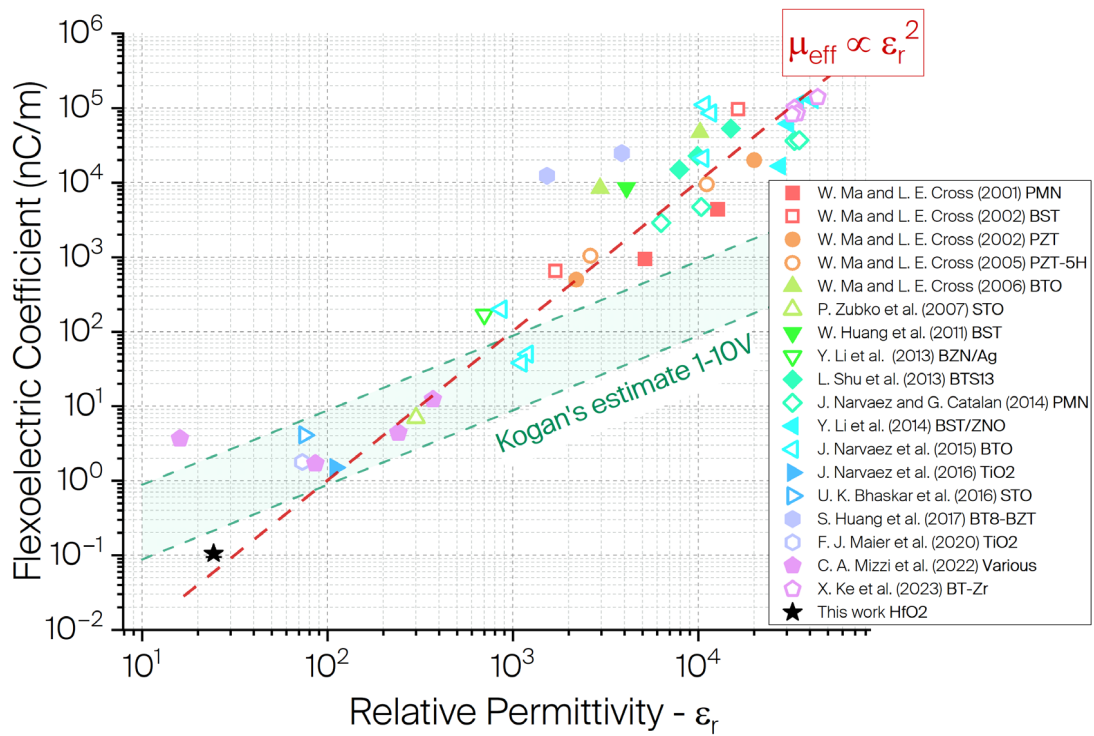

**Figure 6.** Relation between flexoelectric coefficients and relative permittivity: This figure compiles flexoelectric coefficients from various literature sources[4,6–19,30,36] and plots them against the measured relative permittivity of the dielectrics. The data reveals a trend where the flexoelectric coefficient seems to be proportional to the square of the relative permittivity. Prior research, including work by Ma and Cross[11], identified anomalies in the linear relationship of flexoelectric coefficients but did not offer a definitive explanation. Measurements of the relative permittivity for amorphous $HfO_2$ can be found in the supplementary material.

**Conclusions**

In this work, we have addressed the critical challenges in the precise measurement of flexoelectric coefficients at the nanoscale. Our methodology carefully distinguishes flexoelectricity from concurrent effects such as piezoelectricity, electrostriction, and electrostatics. We achieve a detection limit as low as 1 fC/m, enabling the measurement of flexoelectricity in any material.

By applying this methodology to microfabricated cantilevers and clamped-clamped beams, we isolated the flexoelectric effect in amorphous $HfO_2$, determined to be 105 ± 10 pC/m. This finding not only represents the first measurement of flexoelectricity in $HfO_2$ but also in any amorphous material, broadening the scope for using glassy materials in flexoelectric applications.

Additionally, our study provides experimental evidence that the flexoelectric coefficient does not seem to depend linearly on the relative permittivity. A quadratic relationship is observed across numerous literature flexoelectric measurements. This observation may indicate the existence of other effects, possibly including piezoelectric contributions in some cases. For this reason, we recommend the adoption of our methodology to distinctly separate flexoelectric effects.

Future work should focus on extending this measurement technique to a broader range of materials, both amorphous and crystalline, to build a robust catalog of flexoelectric coefficients. Investigating materials of technological interest could pave the way for broader applications of flexoelectricity, notably in the creation of lead-free electromechanical actuators suitable for integration in NEMS.

**Supplementary material section**

The supplementary material covers the microfabrication of cantilevers and clamped-clamped beams, X-ray diffraction and SEM analyses to confirm the amorphous nature and thickness of the $HfO_2$ layer, and P-E loop measurements to determine the material's relative permittivity. Methods for isolating and measuring the flexoelectric coefficient using a Laser Doppler Vibrometer are explained, along with a full derivation of the curvature formula and a comprehensive theoretical expression for the flexoelectric coefficient. Additionally, calculations for determining the detection limit of our method are provided. The material also presents simulations comparing resonance frequency shifts in cantilevers and clamped-clamped beams to elucidate flexoelectric effects.

**Acknowledgments**

The authors thank Gustau Catalan and Dragan Damjanovic for the fruitful discussions on the results. This work was supported by financial support from the Swiss National Science Foundation via grants PP00P2_170590 and CRSII5_189967.

**Author declarations section**

The authors have no conflicts to disclose.

**Daniel Moreno-Garcia:** Conceptualization (equal); Data curation (lead); Formal Analysis (lead); Investigation (lead); Methodology (equal); Software (lead); Validation (lead); Visualization (lead); Writing – original draft (lead); Writing – review & editing (lead). **Kaitlin M. Howell:** Conceptualization (equal); Investigation (equal); Writing – review & editing (equal). **Guillermo Villanueva:** Conceptualization (Lead); Funding acquisition (Lead); Project administration (Lead); Resources (Lead); Supervision (Lead); Writing – review & editing (Equal).

**Data availability statement**

The data that support the findings of this study are available from the corresponding author upon reasonable request.

# SI - Flexoelectricity in Amorphous Hafnium Oxide (HfO2)

*Daniel Moreno-Garcia[1], Kaitlin M. Howell[1], Luis Guillermo Villanueva[1]*

[1] Advanced NEMS Group, École Polytechnique Fédérale de Lausanne (EPFL), Lausanne 1015, Switzerland
E-mail: Guillermo.Villanueva@epfl.ch

## Section 1. Microfabrication of the samples

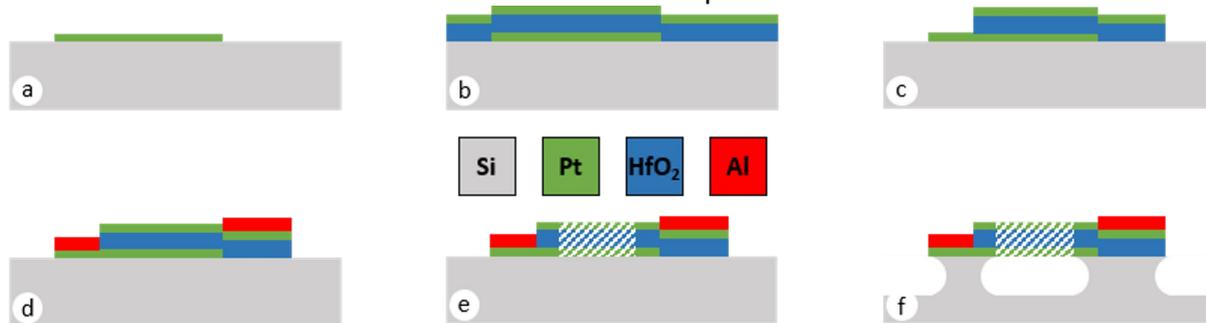

**Figure 1**. Schematic of the main steps to fabricate cantilevers and clamped-clamped beams made of platinum and hafnium oxide. **a)** Liftoff of an evaporated bottom electrode thin film; **b)** ALD deposition of the dielectric and top electrode evaporation; **c)** Top electrode and dielectric patterning and etching; **d)** Aluminum pads liftoff and wafer dicing; **e)** Chip level fabrication to pattern and release the actuators using isotropic Si etching. The checkered area is where the cantilevers are located.

## Section 2. X-Ray Diffraction

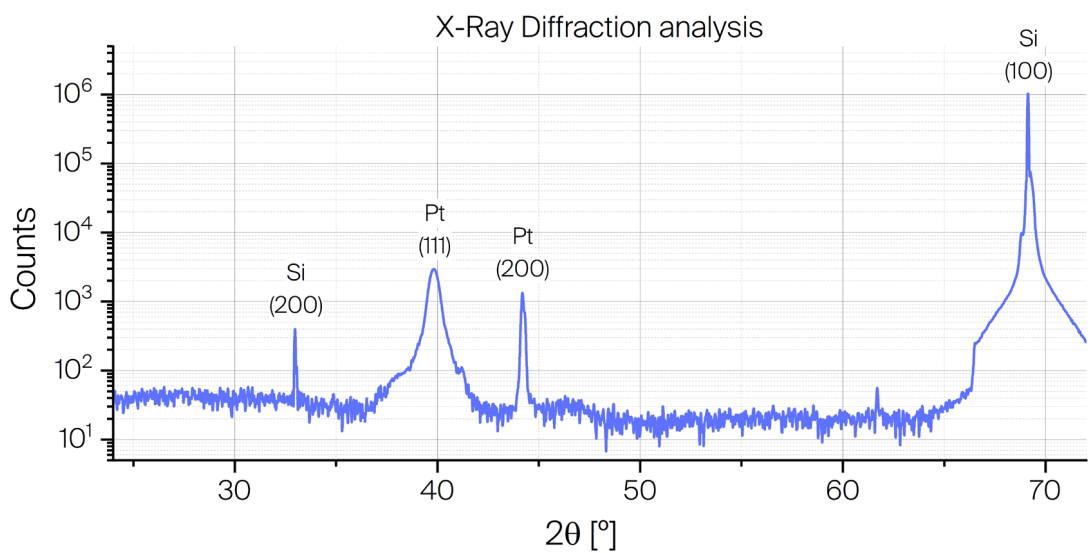

**Figure 2.** X-Ray Diffraction analysis of the samples. We can see the peaks corresponding to the silicon substrate and the platinum of the electrodes. There is no peak from the HfO2, determining that the dielectric is in amorphous form.

# Section 3. Thickness verification with SEM cross-section.

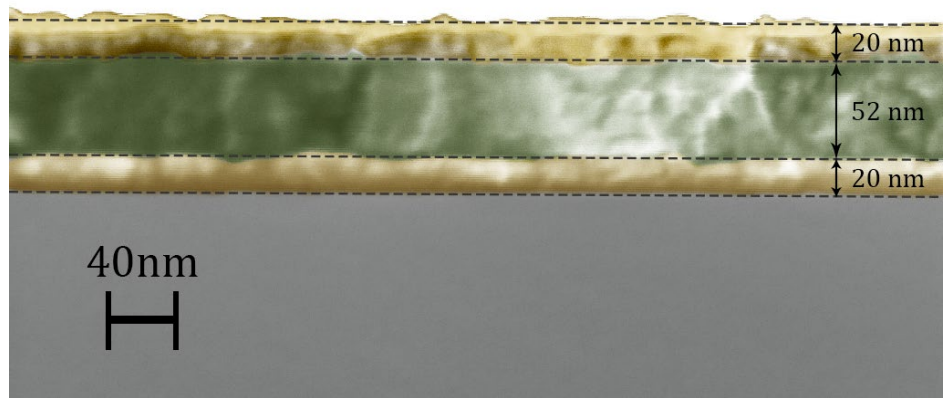

*Figure 7. Cross-section of the layers that compose the cantilevers and clamped-clamped beams. Thanks to this, we could determine that the platinum electrodes measured 20 nm in thickness and the flexoelectric material 52 nm.*

# Section 4. P-E Loop Measurement. Relative permittivity

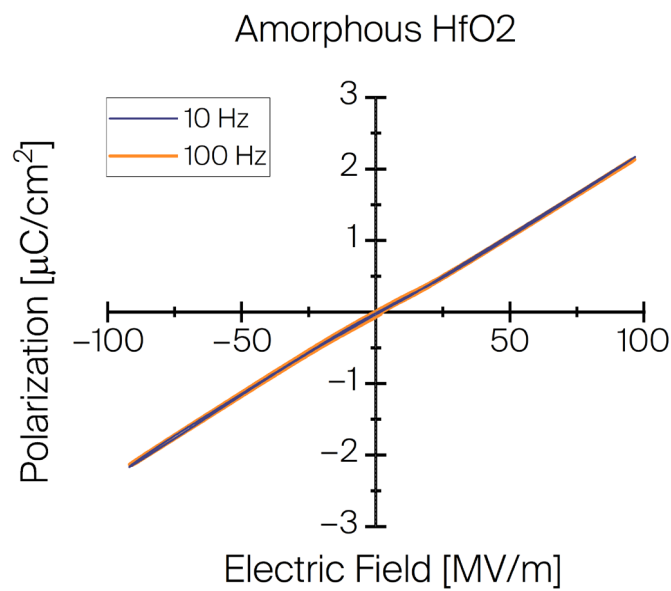

*Figure 8. PE loop measurement (Polarization – Electric Field). This measurement shows that the material is not ferroelectric, as it does not exhibit hysteresis.*

From these measurements, we can also extract the relative permittivity of the material as follows:

$$\varepsilon_r = \frac{P}{\varepsilon_0 E} + 1$$

Applying this formula to several measurements, the obtained relative permittivity for amorphous hafnium oxide is:

$$\varepsilon_r(HfO_2) = 24 \pm 3$$

## Section 5. Input signals to isolate the flexoelectric coefficient.

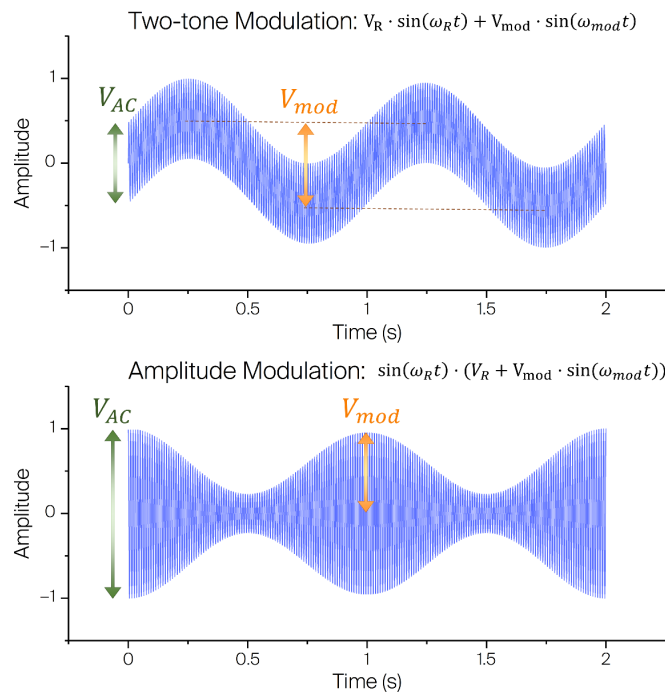

**Figure 3.** *Input signals are used in the cantilevers to isolate the flexoelectric effect. Two-tone modulation helps obtain the contribution of electrostatic and electrostrictive effects. By using Amplitude modulation, we can obtain the contribution of flexoelectricity and piezoelectricity.*

## Section 6. Setup to measure the flexoelectric coefficient.

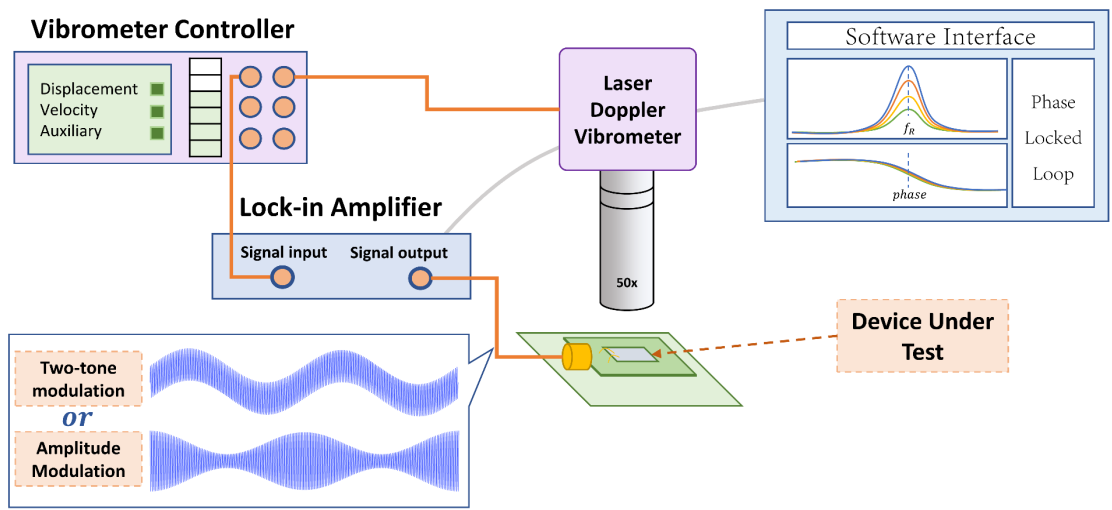

**Figure 4.** *Complete set-up to measure the flexoelectric coefficient of a sample. A Laser Doppler Vibrometer (LDV) points to the cantilever and calculates its displacement. A lock-in amplifier actuates with the corresponding modulations and reads the data from the LDV. This data is demodulated to see the influence of the actuation at the modulation frequency. The Lock-in amplifier also keeps track of the resonance frequency using a Phase Locked Loop (PLL).*

## Section 7. Derivation of the curvature formula

$$\kappa(V) = \frac{\mu_{eff}V}{D_f} + \frac{e_{31}Z_pV}{D_f} - \frac{1}{2}\frac{\nu_d\varepsilon_r\varepsilon_0 Z_P V^2}{t_d D_f} + \frac{2\cdot E_Y M Z_P V^2}{t_d D_f}$$

This formula is derived from the curvature created by the effect of a bending moment on the beam. The relation between the curvature ($\kappa$) and the bending moment ($M_z$) uses the Bending Stiffness $\langle EI_{z,y_0}\rangle$ [1].

$$M_z(x) = \kappa \langle EI_{z,y_0}\rangle$$

The relation between bending stiffness and flexural rigidity is the following[1], where W is the cantilever's width:

$$D_f = \frac{\langle EI_{z,y_0}\rangle}{W}$$

Which leads to a curvature formula as follows:

$$\kappa = \frac{M_z(x)}{D_f \cdot W}$$

### Flexoelectric component

A bending moment is created from the material polarization. The bending moment related to this effect is: $M_z(x) = \mu_{eff} \cdot V \cdot W$. This leads to the curvature formula commonly used for flexoelectricity[2].

$$\kappa_{flexo} = \frac{\mu_{eff}V}{D_f}$$

### Piezoelectric component

The moment that bends the cantilever through piezoelectricity comes from the elongation of the dielectric layer under a voltage. The piezoelectric coefficient $e_{31}$ relates the voltage applied on the electrodes with the elongation of the dielectric. This elongation can create a curvature moment if the center of the cantilever is not aligned with the center of the dielectric ($Z_p \neq 0$).

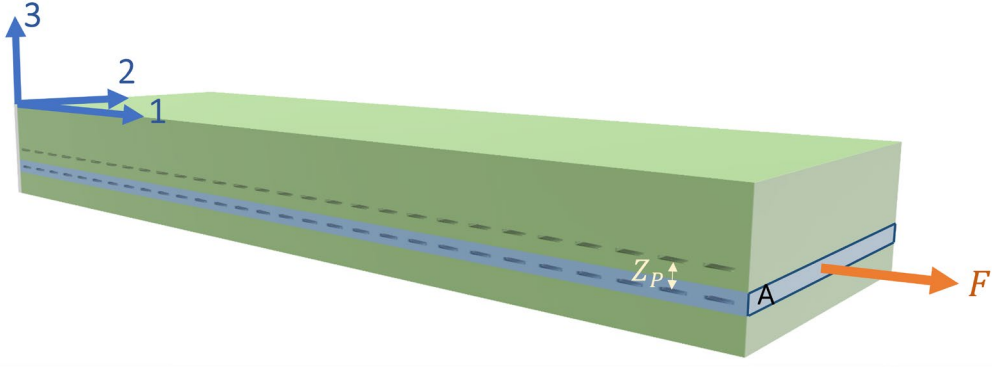

*Figure 5. Schematic of a cantilever to show the definition of the axis, the distance between the center of the cantilever and the neutral axis ($Z_p$) and example of a longitudinal force (F) on the dielectric.*

$$M_z(x) = F_1 \cdot Z_p$$

Here, $F_1$ represents the force resulting from piezoelectric expansion, which can be expressed in terms of the generated piezoelectric stress $\sigma_{11}$ and the dielectric cross-sectional area $A$. This area is the product of the dielectric's width ($W$) and the dielectric's thickness ($t_p$).

$$F_1 = \sigma_{11} \cdot A = \sigma_{11} \cdot t_p \cdot W$$

From the constitutive equations of piezoelectricity[3], the strain $\varepsilon_{11}$ is related to the electric field $E_3$ trough the piezoelectric coefficient $d_{113}$:

$$\varepsilon_{11} = d_{113} \cdot E_3$$

Considering a small perturbation of the dielectric, the strain can be written as stress divided by the Young's modulus. This provides a relationship between $\varepsilon_{31}$ and $\sigma_{11}$:

$$\varepsilon_{11} = \frac{\sigma_{11}}{E_Y} = d_{113} \cdot E_3$$

Substituting in the Force:

$$F = E_Y \cdot d_{113} \cdot E_3 \cdot t_P \cdot W$$

Substituting in the moment and converting the electric field into voltage ($V_3 = E_3 \cdot t_P$):

$$M_z(x) = E_Y \cdot d_{113} \cdot V_3 \cdot W \cdot Z_p$$

Finally, the curvature generated by the piezoelectricity is:

$$\kappa_{piezo} = \frac{E_Y \cdot d_{113} \cdot V_3 \cdot Z_p}{D_f}$$

## Electrostatic component

Similarly to the piezoelectric component, we decompose the moment in a force along the longitudinal axis of the cantilever and the distance between the center of the dielectric and the neutral axis ($Z_p$).

$$M_z(x) = F_1 \cdot Z_p$$

For the electrostatic effect, the two electrodes feel a force ($F_{ES}$, in axis 3) that pushes them closer together. The electrostatic force depends on the relative and vacuum permittivities ($\epsilon_r, \epsilon_o$), the area of the electrodes ($A$), the distance between the electrodes ($d$), the displacement of the plates ($z$) and the applied voltage ($V$).

$$F_{ES} = F_3 = \frac{1}{2}\frac{\partial C}{\partial z}V^2 = \frac{1}{2}\frac{\epsilon_r \epsilon_o A}{(d-z)^2}V^2$$

Doing a Taylor expansion:

$$F_3 = \frac{1}{2}\frac{\epsilon_r \epsilon_o A}{(d-z)^2}V^2 = \frac{1}{2}\frac{\epsilon_r \epsilon_o A}{d^2}V^2\left(1 + 2\frac{z}{d} + 3\left(\frac{z}{d}\right)^2 + \cdots\right) \approx \frac{1}{2}\frac{\epsilon_r \epsilon_o A}{d^2}V^2$$

We make explicit the area A, and $d = t_d$:

$$F_3 = \frac{1}{2}\frac{\epsilon_r \epsilon_o \cdot L \cdot W}{t_d^2}V^2$$

The electrostatic force happens along the z-axis (axis 3), we are interested in how much of this force will be happening in a perpendicular direction x (axis 1). This conversion is made through the Poisson ratio $\nu_d$, which can be defined as the ratio between longitudinal stress ($\sigma_1$) and vertical stress ($\sigma_3$).

$$\nu_d = -\frac{\sigma_1}{\sigma_3}$$

The forces happening on the dielectric are:

$$F_3 = \sigma_3 \cdot W \cdot L$$

$$F_1 = \sigma_1 \cdot t_d \cdot W$$

We do the ratio and substitute for the Poisson ratio:

$$F_1 = -F_3 \cdot \frac{t_d \nu_d}{L} = -\frac{1}{2}\frac{\epsilon_r \epsilon_o \nu_d W}{t_d}V^2$$

The curvature is:

$$\kappa_{electrostatic} = -\frac{1}{2}\frac{\epsilon_r \epsilon_o \nu_d Z_P}{t_d D_f}V^2$$

### Electrostrictive component

In this effect, the existence of a DC voltage (or a very low-frequency voltage, as in our case) polarizes the dielectric material. This makes the material have an apparent piezoelectricity. In this regard, we could talk about an apparent piezoelectric voltage that depends on the electrostrictive coefficient ($M$), which can be defined as[4]:

$$d_{113} = 2 \cdot M \cdot E_3$$

By inserting this apparent piezoelectric voltage in the curvature definition for piezoelectricity, we get:

$$\kappa_{electrostrictive} = \frac{2 \cdot E_Y \cdot M \cdot V_3^2 \cdot Z_p}{t_d \cdot D_f}$$

# Section 8. Detailed calculation of the flexoelectric coefficient

The Laser Doppler Vibrometer (LDV) is used to measure the displacement amplitude at the point targeted by the laser. To ensure accurate measurement of the displacement amplitude at the cantilever's tip, we calibrate using the thermomechanical noise (TMN). This calibration involves recording the thermomechanical noise at the resonance frequency and comparing it with the theoretically expected value. Such comparison enables us to accurately convert the measured displacement at any point along the cantilever into the displacement at the tip.

The theoretical Thermomechanical Noise at resonance ($S_x(\omega_0)$)[1], will depend on the quality factor (Q), Boltzmann constant ($k_B$), the temperature (T), the effective mass of the cantilever at for the first mode ($m_{eff}$) and the resonance frequency ($\omega_0$).

$$S_x(\omega_0) = \frac{4 \cdot Q \cdot k_B \cdot T}{m_{eff} \cdot \omega_0^3}; \quad in \; [m^2/Hz] \tag{1}$$

The measured TMN will be represented in Power Spectral Density form (PSD($\omega$)), with units of $V^2/Hz$). We define a calibration factor ($\alpha$) that when multiplied by the measured PSD, yields the theoretically expected thermomechanical noise. This calibration factor will convert the measured volts in the lock-in amplifier ($V_{LDV}$) into displacement at the tip of the cantilever.

$$S_x(\omega_0) = \alpha^2 \cdot PSD(\omega_0) \tag{2}$$

$$\alpha = \sqrt{\frac{4 \cdot Q \cdot k_B \cdot T}{m_{eff} \cdot \omega_0^3 \cdot PSD}}; \quad in \; [m/V] \tag{3}$$

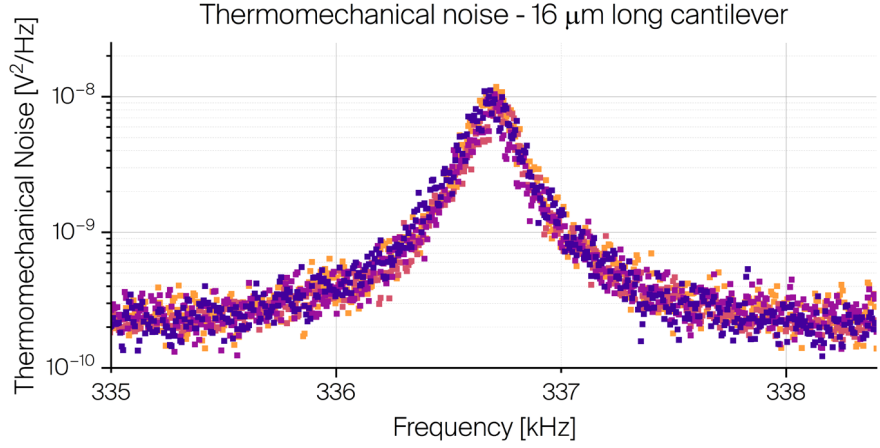

*Figure 5.* *Thermomechanical noise (TMN) measurement of a 16 μm long cantilever. The measurements are performed by shining the Laser Doppler Vibrometer on the cantilever and reading the noise of the signal, without any kind of actuation. This measurement, when compared to the expected theoretical TMN of the cantilever, helps us translate a measurement on any location of the beam into the displacement of the tip of the cantilever.*

Using this, the experimental displacement at the tip at the resonance frequency of the cantilever will be given by $u_n(\omega_n)$.

$$u_n(\omega_n) = V_{LDV} \cdot \sqrt{\frac{4 \cdot Q \cdot k_B \cdot T}{m_{eff} \cdot \omega_0^3 \cdot PSD}} \qquad (4)$$

The effective mass for a cantilever on the first mode of vibration is given by $m_{eff}$, which includes the mass of the cantilever, calculated as the geometrical volume (length – L, width – W, and thickness – t) times the density of each material ($\rho_i$).

$$m_{eff} = \frac{1}{4} L \cdot W \cdot \Sigma \rho_i t_i \qquad (5)$$

$$u_n(\omega_n) = V_{LDV} \cdot \sqrt{\frac{4 \cdot Q \cdot k_B \cdot T}{\frac{1}{4} L \cdot W \cdot \Sigma \rho_i t_i \cdot \omega_0^3 \cdot PSD}} \qquad (6)$$

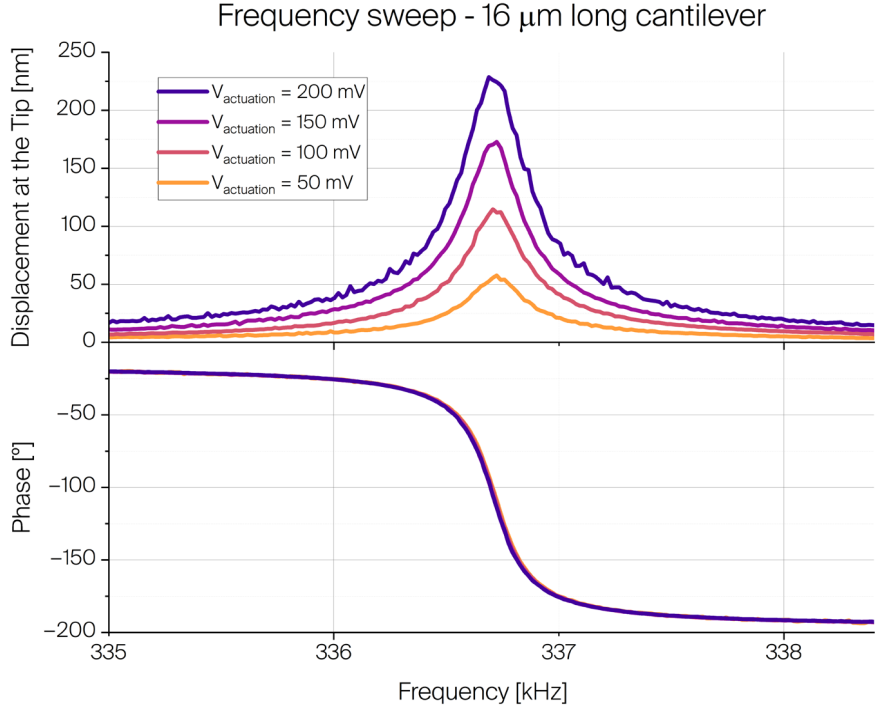

**Figure 6.** *Displacement at the tip of a 16 µm long cantilever when actuated with an AC signal. The measured amplitude in Volts provided by the LDV has been converted into nm movement at the tip of the cantilever using Equation 6.*

To relate the displacement at the cantilever's tip with the flexoelectric coefficient, we refer to established literature[1] that provides the relationship between the amplitude of the tip displacement ($u_n(\omega_n)$) due to an applied moment ($M(\omega)$). This relationship is influenced by the flexural rigidity ($D_f$) and a constant that varies based on the mode of vibration and electrode coverage ($\chi_n^A$). Specifically, for the first mode of a cantilever with full electrode coverage, the constant $\chi_n^A \approx 0.445$, as detailed in the reference[1].

$$u_n(\omega_n) \approx \chi_n^A \cdot \frac{M(\omega) \cdot L^2}{D_f \cdot W} \cdot Q \tag{7}$$

The moment generated by the flexoelectric effect $M_{flexo}$ will be linear depending on the actuation voltage ($V_{actuation}$).

$$M_{flexo} = \mu_{eff} \cdot V_{actuation} \cdot W \tag{8}$$

$$u_n(\omega_n) \approx 0.445 \cdot \frac{\mu_{eff} \cdot V_{actuation} \cdot L^2}{D_f} \cdot Q \tag{9}$$

Using equation (4) and (9):

$$\mu_{eff} = \frac{V_{LDV}}{0.445 \cdot V_{actuation} \cdot L^2 Q} \cdot \sqrt{\frac{4Q k_B T}{\frac{1}{4} L \cdot W \cdot \Sigma \rho_i t_i \cdot \omega_0^3 \cdot PSD}} \cdot D_f \tag{10}$$

The flexural rigidity ($D_f$) for our case with symmetrical top-bottom electrodes and full coverage is calculated as:

$$D_f = E_{HfO2} \frac{t_{HfO2}^3}{12} + E_{Pt} \frac{1}{12}\left[(2t_{Pt} + t_{HfO2})^3 - t_{HfO2}^3\right] \qquad (11)$$

Arranging everything we obtained the complete expression.

$$\mu_{eff} = \frac{V_{LDV}}{0.445 \cdot V_{actuation} \cdot L^2 Q} \cdot \sqrt{\frac{4Qk_BT}{\frac{1}{4}L \cdot W \cdot \Sigma\rho_i t_i \cdot \omega_0^3 \cdot PSD}} \cdot \left[E_{HfO2} \frac{t_{HfO2}^3}{12} + E_{Pt} \frac{1}{12}\left[(2t_{Pt} + t_{HfO2})^3 - t_{HfO2}^3\right]\right] \qquad (12)$$

$$\mu_{eff} = \underbrace{V_{LDV} \cdot \sqrt{\frac{4Qk_BT}{\frac{1}{4}L \cdot W \cdot \Sigma\rho_i t_i \cdot \omega_0^3 \cdot PSD}} \cdot \frac{1}{0.445 \cdot L^2 Q}}_{\kappa_{flexo}} \cdot \underbrace{\frac{\left[E_{HfO2} \frac{t_{HfO2}^3}{12} + E_{Pt} \frac{1}{12}\left[(2t_{Pt} + t_{HfO2})^3 - t_{HfO2}^3\right]\right]}{V_{actuation}}}_{D_f}$$

Output from LDV | Translate the amplitude to the tip of the cantilever | Tip displacement to beam curvature

This formula is automatically applied for every measurement, with the variables $V_{LDV}$, $PSD$, $Q$ and $\omega_0$ are calculated in real-rime for every $V_{actuation}$ that we apply to the cantilevers. The geometries of the cantilevers are measured using SEM, while the thickness of the layers is determined with the Filmetrics F54.

To determine the Young's modulus, we measured the resonance frequencies of our amorphous cantilevers and fit these to simulations, obtaining the Young's modules with a 10% uncertainty. All values are measured at room temperature.

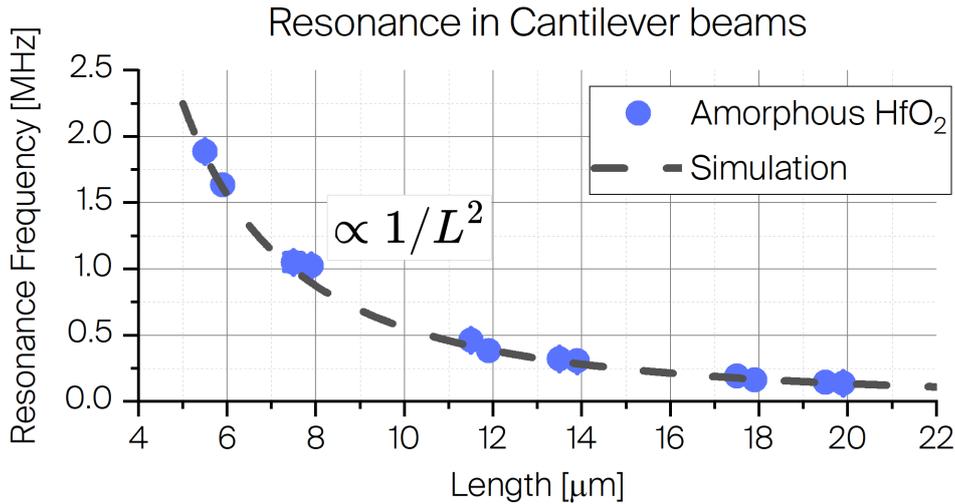

**Figure 7.** *Measured resonance frequencies for cantilevers of varying lengths. The finite-element simulation allows us to obtain the effective Young's modulus for the materials.*

A summary of the variables used in our measurements is provided in the following table:

| Variable | Method | Value range | Uncertainty |
|---|---|---|---|
| $V_{LDV}$ | Output from LDV | $\sim 100\ mV$ | 0.1% |
| $Q$ |  | $\sim 1500$ | 1% |
| $f_0$ |  | $100\ kHz - 2\ MHz$ | 0.01% |
| $PSD$ |  | $\sim 10^{-8}\ V^2/Hz$ | 1% |
| $L$ | Scanning Electron Microscope | $5.9\ nm - 19.9\ nm$ | 2% |
| $W$ |  | $2.27\ nm$ | 2% |
| $k_B T$ | Typical value @ 298 K | $4.11 \cdot 10^{-21}\ J$ | 2% |
| $\rho$ | Cleanroom database | $21450\ /\ 9680\ kg/m^3$ | - |
| $t$ | Filmetrics F54 | $20\ nm\ /\ 52\ nm$ | 0.2% |
| $E_{HfO2}$ | Resonance + Simulation | $170\ GPa$ | $\sim 2\%$ |
| $E_{Pt}$ |  | $200\ GPa$ | $\sim 2\%$ |
| $V_{actuation}$ | Input voltage | $\sim 100\ mV$ | - |

## Section 9. Limit of detection of our method.

The presented method is limited by thermomechanical noise. This means that we can distinguish the minimum cantilever movement by flexoelectricity until this is of the same level as the thermomechanical noise.

The thermomechanical noise has the following formula[1]:

$$S_x(\omega_0) = \frac{4 \cdot Q \cdot k_B \cdot T}{m_{eff} \cdot \omega_0^3}; \qquad [m^2/Hz]$$

The movement of the tip of the cantilever is given by the flexoelectric curvature:

$$w(L) = 0.445 \cdot Q \cdot L^2 \cdot \kappa = 0.445 \cdot Q \cdot L^2 \cdot \frac{\mu_{eff}}{D_f} \cdot V$$

Equalizing the thermomechanical noise expression and the movement of the cantilever tip:

$$\sqrt{\frac{4 \cdot Q \cdot k_B \cdot T}{m_{eff} \cdot \omega_0^3} \cdot BW} = 0.445 \cdot Q \cdot L^2 \cdot \frac{\mu_{eff}}{D_f} \cdot V$$

Using a simplified expression for the flexural rigidity for a low Width-to-Length ratio:

$$D_f \approx \frac{E}{12} t_{TOT}^3$$

We can isolate the minimum flexoelectric coefficient that we can detect:

$$\mu_{eff} = \frac{E}{12} t_{TOT}^3 \cdot \frac{1}{V \cdot 0.445 \cdot Q \cdot L^2} \sqrt{\frac{4 \cdot Q \cdot k_B \cdot T}{m_{eff} \cdot \omega_0^3} \cdot BW}$$

Where:

| $E$: Young's modulus | $Q$: Quality factor | $k_B$: Boltzmann constant |
| --- | --- | --- |
| $\omega_0$: Resonance Freq | $V$: Applied Voltage | $T$: Temperature |
| $t_{TOT}$: Total Thickness | $L$: Cant. length | $m_{eff}$: Effective mass |
| $BW$: Bandwidth Meas | | |

The evaluation of this formula gives us a minimum detectable flexoelectric coefficient around $1\ fC/m$.

## Section 10. COMSOL Simulations.

Comparison between the resonance frequency shifts generated in Cantilevers and Clamped-clamped beams. These simulations assume that piezoelectricity is present in the devices and obtain the increase of resonance frequency shift in the clamped-clamped devices, compared to the cantilevers.

a) Cantilever – In hertz units

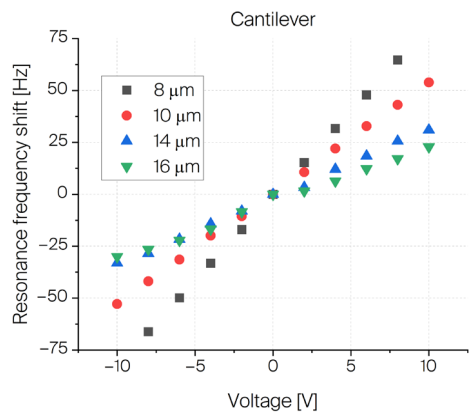

b) Clamped-Clamped – In hertz units

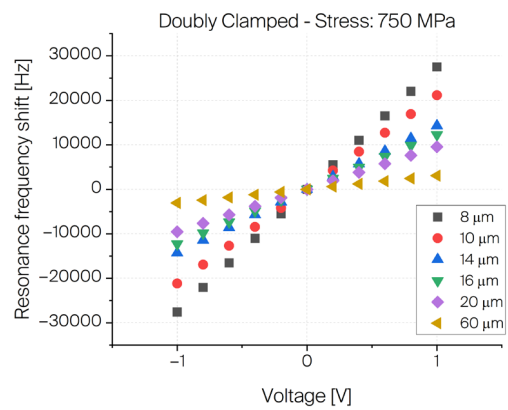

c) Cantilever – In relative units

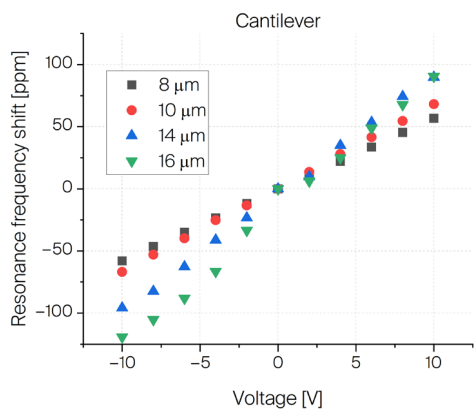

d) Clamped-Clamped – In relative units

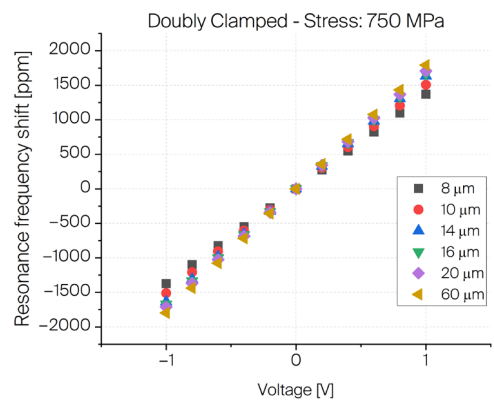

**Figure 6.** *Here we are simulating cantilevers and clamped-clamped beams with 3 um widths and a varying length, assuming piezoelectricity in the samples. The simulations include the internal stress measured in the layers (750 MPa) after depositions in the cleanroom. a) and c) show the same information in hertz units and relative units (part per million, ppm). The same with b) and d) for the case of the clamped-clamped beams. We can observe in c) that the application of a voltage on the cantilevers produces a change in the resonance frequency of 90 ppm for 10 V (9 ppm/V). d) Shows the analogous case for the clamped-clamped devices, where 1750 ppm are achieved for an actuation of 1 V (1750 ppm/V). This means that the clamped-clamped devices show a relative change in the resonance frequency of around 200 times larger than the cantilever for the same actuation.*

The following simulations have allowed us to understand how flexoelectricity can produce changes in resonance frequency for cantilevers and clamped-clamped devices. To simulate the flexoelectric effect we have created a bimorph in which one layer includes stress in X and Y, providing a curvature in both directions.

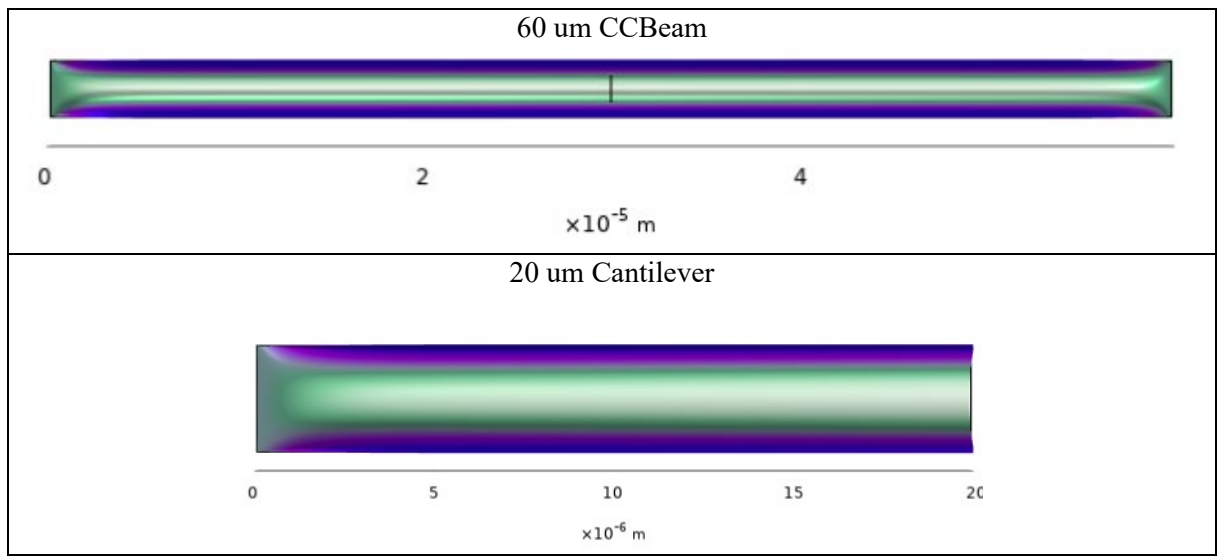

**Figure 7.** *Images depicting the lateral bending caused by flexoelectricity in cantilevers and clamped-clamped beams. Simulations indicate that flexoelectricity can induce similar shifts in resonance frequency of both types of devices. For this effect to happen at the modulation voltage and to be detected using the two-tone modulation, an initial lateral curvature offset of tens of nanometers is necessary. Nonetheless, accurately measuring the actual lateral curvature of the devices has proven to be challenging.*